\documentclass[12pt,a4paper]{article}
\usepackage{amsmath}
\usepackage{graphicx}
\usepackage{epsfig}
\newcommand{\be}{\begin{equation}}
\newcommand{\ee}{\end{equation}}
\newcommand{\ben}{\begin{equation*}}
\newcommand{\een}{\end{equation*}}
\newcommand{\bea}{\begin{eqnarray}}
\newcommand{\eea}{\end{eqnarray}}
\newcommand{\ar}{\begin{array}}
\newcommand{\arn}{\end{array}}

\newcommand{\vk}{\vec{k}}
\newcommand{\vks}{\vec{k}^{\;2}}
\newcommand{\q}{\vec{q}}
\newcommand{\qs}{\vec{q}^{\;2}}
\newcommand{\qp}{\vec{q}^{\;\prime}}
\newcommand{\qps}{\vec{q}^{\;\prime\; 2}}
\newcommand{\p}{\vec{p}}
\newcommand{\ps}{\vec{p}^{\;2}}

\newcommand{\x}{\vec{r}}

\def\pnot{\mbox{${\not{\hbox{\kern-3.0pt$p$}}}$}}
\def\qnot{\mbox{${\not{\hbox{\kern-2.0pt$q$}}}$}}
\def\enot{\mbox{${\not{\hbox{\kern-2.0pt$e$}}}$}}
\def\knot{\mbox{${\not{\hbox{\kern-2.0pt$k$}}}$}}

\def\fun#1#2{\lower3.6pt\vbox{\baselineskip0pt\lineskip.9pt\ialign
{$\mathsurround=0pt#1\hfil##\hfil$\crcr#2\crcr\sim\crcr}}}

\begin{document}

\textwidth=135mm
 \textheight=200mm

\begin{titlepage}

\hskip 11cm \vbox{ \hbox{Budker INP 2013-6}  }
\vskip 3cm
\begin{center}
{\bf M\"{o}bius invariant BFKL equation for  the  adjoint representation  in
$N=4$ SUSY}
\end{center}

\vskip 0.5cm

\centerline{V.S.~Fadin$^{a\,\dag}$, R.~Fiore$^{b\,\ddag}$, L.N.~Lipatov$^{c\,\S}$,
A.~Papa$^{b\,\P}$}

\vskip .6cm

\centerline{\sl $^{a}$ Budker Institute of Nuclear Physics of SD RAS, 630090 Novosibirsk
Russia}
\centerline{\sl and Novosibirsk State University, 630090 Novosibirsk, Russia}
\centerline{\sl $^{b}$ Dipartimento di Fisica, Universit\`a della Calabria,}
\centerline{\sl and Istituto Nazionale di Fisica Nucleare, Gruppo collegato di
Cosenza,}
\centerline{\sl Arcavacata di Rende, I-87036 Cosenza, Italy}
\centerline{\sl $^{c}$ Petersburg Nuclear Physics Institute}
\centerline{\sl and St. Petersburg State University,
Gatchina, 188300, St. Petersburg, Russia}

\vskip 2cm

\begin{abstract}

It is shown that in the next-to-leading approximation of $N=4$ SUSY the BFKL equation
for two-gluon composite states in the adjoint representation of the gauge group can be
reduced to a form which is invariant under M\"{o}bius transformation in the
momentum space. The corresponding similarity transformation of its integral kernel is
constructed in an explicit way.
\end{abstract}


\vfill \hrule \vskip.3cm \noindent $^{\ast}${\it Work supported 
in part by the Ministry of Education and Science of Russian Federation,
in part by  RFBR,  grant 13-02-01023, and in part by Ministero Italiano dell'Istruzione,
dell'Universit\`a e della Ricerca.} \vfill $
\begin{array}{ll} ^{\dag}\mbox{{\it e-mail address:}} &
\mbox{fadin@inp.nsk.su}\\
^{\ddag}\mbox{{\it e-mail address:}} &
\mbox{roberto.fiore@cs.infn.it}\\
^{\S}\mbox{{\it e-mail address:}} &
\mbox{lipatov@thd.pnpi.spb.ru}\\
^{\P}\mbox{{\it e-mail address:}} &
\mbox{alessandro.papa@cs.infn.it}\\
\end{array}
$

\end{titlepage}

\vfill \eject

\section{Introduction}

The BFKL equation for the Pomeron wave function in the color singlet representation is
well known~\cite{BFKL}. In particular for the description of total cross-sections at
high energies $\sqrt{s}$ its simple form at vanishing momentum transfers
$|q|=\sqrt{-t}=0$ is used. The integral kernel of this equation was calculated in the
next-to-leading order (NLO) for QCD ~\cite{Fadin:1998py} and for $N=4$
SUSY~\cite{Kotikov:2000pm}. But the BFKL approach is applicable also for arbitrary
$t$-channel color states constructed from two gluons. The corresponding NLO kernels at
momentum transfers $q\ne 0$ are known both in QCD~\cite{Fadin:1998jv} and in $N=4$
SUSY~\cite{Fadin:2007xy}. For the phenomenological applications the most important cases
are the color singlet states constructed from two or several reggeized gluons. The
corresponding Regge poles appear in the amplitudes having the antisymmetric adjoint
representation (the $f$-coupling) in the $t$-channel. The concept of gluon
reggeization was formulated on the base of the fixed-order calculations~\cite{BFKL} and
was checked in the leading logarithmic approximation (LLA) with the use of the
so-called bootstrap equations~\cite{Balitskii:1979}, which follow from the
compatibility of the multi-Regge form of production amplitudes with the $s$-channel
unitarity.
Later the bootstrap equations were constructed in the NLO~\cite{Fadin:2006bj}.
Now the fulfillment of the corresponding relations in the NLO is proved
(see~\cite{Kozlov:2012zza} and references therein). Note that one can use the effective
action for the calculation of the gluon Regge trajectory and the reggeon couplings in
upper orders of perturbation theory~\cite{eff}.

There are at least two other reasons for the significance of the BFKL kernel  in
the adjoint representation.
Its first application is related to the Bartels-Kwiecinski-Praszalowicz (BKP)
equation~\cite{Bartels:1980pe}. This equation describes bound states of several
reggei\-zed gluons, in particular the Odderon which is a C-odd three-gluon state. In the
last case, the pair-wise part of the NLO BKP kernel contains the NLO BFKL kernel for the
symmetric adjoint representation (the $d$-coupling)~\cite{Bartels:2012sw}.
Note that the difference between the symmetric and anti-symmetric representations
appears only in NLO and even in this case the corresponding kernels coincide in the
limit of large number of colors or provided  that all particles in the action
belong to the adjoint representation of the gauge group, as in the $N=4$ SUSY.
Another application of the BFKL approach was suggested in the framework of $N=4$ SUSY to
verify and generalize the Bern-Dixon-Smirnov (BDS) ansatz~\cite{Bern:2005iz} for the
production amplitudes with maximal helicity violation in the limit of large number of
colors. It gave a possibility to find the high-energy behavior of the remainder
function for the BDS ansatz~\cite{Bartels:2008ce,Lipatov:2010qg,Bartels:2010tx} and to
establish the relation of this problem with an integrable open spin
chain~\cite{Lipintopen}.

The remainder function for the 6-point scattering amplitude in the kinematical regions
containing the Mandelstam cut contribution was calculated recently in
NLO~\cite{Fadin:2011we}, reproducing the three-loop expression for this function suggested
by L.~Dixon and collaborators~\cite{Dixon:2011pw}.
It was done using the NLO BFKL kernel for the adjoint
representation of the gauge group with subtracted gluon trajectory depending on the
total momentum transfer $\q$. The eigenvalues of the kernel at large $\qs$ were found
and the BFKL equation was solved assuming the M\"{o}bius invariance of the modified
kernel $\hat{\cal K}$ (with an omitted factor $\qs$) in the two-dimensional transverse
momentum space. The existence
of the M\"{o}bius invariant kernel $\hat{\cal K}$ follows from general arguments related
to the dual conformal invariance of the remainder function. However, the known NLO
expression for $\hat{\cal K}$ obtained in the standard approach is not conformal
invariant. In principle, this  does not contradict the above assumption, because
the NLO kernel is scheme-dependent. But for the verification of our assumption a
similarity transformation reducing the standard kernel to the M\"{o}bius invariant
expression should exist. Below we construct such transformation in the momentum space
explicitly.

\section{Standard and M\"{o}bius invariant forms of the kernel}

The modified BFKL kernel $\hat{\cal K}$ for the two-gluon composite state in the
antisymmetric adjoint representation is obtained by subtracting the gluon trajectory
depending on $\q$ and extracting the factor $\qs$ from its initial form:
\begin{equation}
\qs\;K(\vec q_1, \vec q_1^{\;\prime}; \vec q)= \delta^{(2)}(\vec q_1-\vec q_1^{\;\prime})
\vec q_1^{\;2}\vec q_2^{\;2}\left( \omega_{g}(-\vec q_1^{\;2})+\omega_g(-\vec q_2^{\;2})-
\omega_g(-\vec q^{\;2})\right) +
K_r(\vec q_1,\vec q_1^{\;\prime}; \vec q)\,, \label{q2 modified kernel}
\end{equation}
where $\vec{q}_i$ and  $\vec{q}_i^{\;\prime}$ for $i=1,2$ are two reggeon momenta,
$\vec{q}= \vec{q}_1+\vec{q}_2 = \vec{q}_1^{\;\prime}+\vec{q}_2^{\;\prime},
\;\;$ $\omega _{g}$ is the trajectory and $K_r$ is the contribution coming from real
particle production. The term $\omega_g(-\vec q^{\;2})$ is subtracted because the
modified kernel is used for finding the high-energy behavior of the conformal invariant
remainder function to the BDS ansatz containing the corresponding Regge factor.

Taken separately, the trajectories and the real part are infrared-divergent. It is known
that the divergences are canceled in the singlet (Pomeron) kernel.  It occurs that they
are canceled also in Eq.~(\ref{q2 modified kernel}), just due to the subtraction of
$\omega_g(-\vec q^{\;2})$. The cancelation takes place because of  two important
properties:
first, the singular terms of the trajectories do not depend on reggeon momenta, and
second, the singular contribution to the real part of the kernel for the adjoint
representation is two times smaller than the similar contribution to the Pomeron kernel.

Therefore, the modified kernel can be written in the form (cf.~\cite{Fadin:2011we})
\begin{equation}
K(\vec q_1, \vec q_1^{\;\prime}; \vec q)= \frac{\vec q_1^{\;2}\vec q_2^{\;2}}{2\qs}
\delta^{(2)}(\vec q_1-\vec q_1^{\;\prime})
\left(\omega _g(-\vec q_1^{\;2})+\omega_g(-\vec q_2^{\;2})-
2\omega_g(-\vec q^{\;2})\right) +
K^{f}(\vec q_1,\vec q_1^{\;\prime}; \vec q), \label{K through Kf}
\end{equation}
where both terms are infrared-finite. In particular,
\be
\omega_g(-\vec q_1^{\;2})+\omega_g(-\vec q_2^{\;2})-
2\omega_g(-\vec q^{\;2})= - \frac{\alpha_s \,N_{c}}{2\pi}
\left(1-\zeta (2)\, \frac{\alpha_s  \,N_{c}}{2\pi}
\right)\,\ln\left(\frac{\vec{q}_1^{\;2}\vec{q}_2^{\;2}}{\vec{q}^{\;4}}\right)\,.
\label{difference of omega}
\ee
Note that for the gauge coupling constant we use the dimensional reduction instead of
the dimensional regularization which violates supersymmetry. This corresponds to the
finite charge renormalization
\begin{equation}
\alpha_s(\mu)\rightarrow
\alpha_s(\mu)\left(1-\frac{\alpha_s(\mu)N_c}{12\pi}\right)\,.
\label{finite renormalization}
\end{equation}
One can write $\alpha_s$ instead of $\alpha_s(\mu)$, because in $N=4$ the coupling
constant is not running. Using the results of~\cite{Bartels:2012sw} for the contribution
$K^{f}$ and an integral representation for the difference of the trajectories obtained
in Eq.~(\ref{difference of omega}), we present the modified kernel as follows:
\[
K(\vec q_1, \vec q_1^{\;\prime}; \vec q)=K^{B}(\vec q_1, \vec q_1^{\;\prime}; \vec q)
\left(1-\frac{\alpha_s N_c}{2\pi}\zeta(2)\right)
+\delta^{(2)}(\vec{q}_{1}-\vec{q}^{\;\prime}_{1})
\frac{\vec{q}_1^{\;2}\vec{q}_2^{\;2}}{\qs}
\frac{\alpha_s^2 \,N^2_{c}}{4\pi^{2}}3\zeta(3)
\]
\be
+\frac{\alpha_s^2 \,N^2_{c}}{32\pi^{3}}R(\vec q_1, \vec q_1^{\;\prime}; \vec q)\,,
\label{modified kernel}
\ee
where $K^{B}$ is the leading order kernel. It can be written in a form which is
explicitly conformal invariant,
\[
K^{B}(\vec q_1,\vec q_1^{\;\prime};\vec q)=-\delta^{(2)}(\vec{q}_{1}
-\vec{q}^{\;\prime}_{1})
\frac{\vec{q}_1^{\;2}\vec{q}_2^{\;2}}{\qs}
\frac{\alpha_s \,N_{c}}{4\pi^{2}}\!\int\!\frac{\qs\,d^{2}l}{(\q_1 -\vec{l})^2(\q_2
+\vec{l})^2}
\]
\be
\times\left(
\frac{\qs_1 (\q_2 +\vec{l})^2+\qs_2 (\q_1 -\vec{l})^2}{\qs\vec{l}^{\;2}}-1\right)
+\frac{\alpha_s \,N_{c}}{4\pi^{2}}
\left(  \frac{\vec{q}_{1}^{\;2}\vec
{q}_{2}^{\;\prime\;2}+\vec{q}_{1}^{\;\prime\;2}\vec{q}_{2}^{\;2}}{
\vec{q}^{\;2}\vec{k}^{\;2}}-1\right)\,.
\label{K B}
\ee
Furthermore,  the last term in the kernel~(\ref{modified kernel}) is written as
\[
R(\vec q_1,\vec q_1^{\;\prime},\vec q) =
\frac{1}{2}\left(\ln\left(  \frac{\vec{q}_{1}^{\;2}}{\vec{q}^{\;2}
}\right)  \ln\left(  \frac{\vec{q}_{2}^{\;2}}{\vec{q}^{\;2}}\right)
+\ln\left(  \frac{\vec{q}_{1}^{\;\prime\;2}}{\vec{q}^{\;2}}\right)
\ln\left(  \frac{\vec{q}_{2}^{\;\prime2}}{\vec{q}^{\;2}}\right)
+\ln^{2}\left(  \frac{\vec{q}_{1}^{\;2}}{\vec{q}
_{1}^{\;\prime\;2}}\right)  \right)
\]
\[
 -\frac{\vec{q}_{1}^{\;2}\vec{q}
_{2}^{\;\prime\;2}+\vec{q}_{2}^{\;2}\vec{q}_{1}^{\;\prime\;2}}{\qs\vec{k}^{\;2}
}\ln^{2}\left(  \frac{\vec{q}_{1}^{\;2}}{\vec{q}_{1}^{\;\prime\;2}}\right)
-\frac{\vec{q}_{1}^{\;2}\vec{q}_{2}^{\;\prime\;2}-\vec{q}_{2}^{\;2}\vec{q}
_{1}^{\;\prime\;2}}{2\qs\vec{k}^{\;2}}\ln\left(  \frac{\vec{q}_{1}^{\;2}}{\vec
{q}_{1}^{\;\prime\;2}}\right)\ln\left(
\frac{\vec{q}_{1}^{\;2}\vec{q}_{1}^{\;\prime\;2}}{\vec{k}^{\;4}}\right)
\]
\begin{equation}
+4\frac{(\vec k\times \vec q_1)}{\qs\vec{k}^{\;2}}\left(\vec{k}^{\;2}(\vec q_1
\times \vec q_2)-\vec{q_1}^{\;2}(\vec k \times \vec q_2)-\vec{q_2}^{\;2}(\vec k
\times \vec q_1)\right)I_{\vec q_1,-\vec k}
+\left(  \vec{q}_{1}\leftrightarrow -\vec{q}_{2},\;\;\vec{q}_{1}^{\;\prime
}\leftrightarrow - \vec{q}_{2}^{\;\prime}\right)  ~.  \label{R standard}
\end{equation}
Here $\vec{k}=\vec{q}_{1}-\vec{q}_{1}^{\;\prime}=\vec{q}_{2}^{\;\prime}-\vec{q}_{2},
\;$ $(\vec a\times\vec b)= a_x b_y -a_y b_x$ and
\begin{equation}
I_{\vec p, \vec q}=
\int_{0}^{1}\frac{dx}{(\vec p +x\vec q)^{2}}\ln\left(\frac{\vec p^{\;2}}
{x^2\vec q^{\;2}}\right)~. \label{I p q 1}
\end{equation}
This quantity has the symmetry properties
\begin{equation}
I_{\vec p,\vec q}=I_{-\vec p,-\vec q}=I_{\vec q, \vec p}=I_{\vec p,-\vec p
-\vec q}~,
\label{I definition}
\end{equation}
which are evident from the representation
\begin{equation}
I_{\vec p,\vec q}=\int_{0}^{1}\int_{0}^{1}\int_{0}^{1}\frac{dx_{1}dx_{2}dx_{3}
\delta(1-x_{1}-x_{2}-x_{3})}{(\vec p^{\;2}x_{1}+\vec q^{\;2}x_{2}+(\vec p
+\vec q)^{2}x_{3})(x_{1}x_{2}+x_{1}x_{3}+x_{2}x_{3})}~.\label{I symmetric}
\end{equation}
Other useful representations are
\[
I_{\vec p,\vec q}=\int_{0}^{1}
\frac{dx}{\ps(1-x)+\qs x-(\p+\q)^2 x(1-x)}\ln\left(\frac{\ps(1-x)+\qs x}
{(\p+\q)^2x(1-x)}\right)
\]
\begin{equation}
=\int_{0}^{1}dx\int_{0}^{1}{dz}\;\frac{1}{(\p+\q)^2x(1-x)z+(\qs(1-x)+\ps x)(1-z)}~.
\label{integral I}
\end{equation}
In the kernel~(\ref{modified kernel}) the first two terms are conformal invariant (we
remind that in our normalization the integration measure $\; \qs d\vk/(\qps_1\qps_2)$
is M\"{o}bius invariant), but the contribution $R(\vec q_1,\vec q_1^{\;\prime},\vec q)$
violates this invariance. In~\cite{Fadin:2011we} it was assumed that there is a
conformal invariant representation of the kernel. Since its eigenvalues do not depend on
the representation and on the total momentum transfer, they were found using the limit
\be
|\q_1|\sim |\qp_1|\ll |\q|\approx |\q_2|\approx |\qp_2|\,.
\ee
In this limit the kernel~(\ref{modified kernel}) can be written as
\be
K(z)=K^{B}(z)\left(1-\frac{\alpha_s N_c}{2\pi}\zeta(2)\right)+\delta^{(2)}(1-z)
\frac{\alpha_s^2 \,N^2_{c}}{4\pi^{2}}3\zeta(3)
+\frac{\alpha_s^2 \,N^2_{c}}{32\pi^{3}}R(z)\,,
\label{K(z)}
\ee
where $z=q_1/q'_1$,
\[
 K^B(z) =  \frac{\alpha_sN_c}{8\pi^2}\left(\frac{z+z^*}{|1-z|^2}
-\delta^{(2)}(1-z)\int\frac{d\vec l}{|l|^2}\frac{l+l^*}{|1-l|^2}\right)\,,\;\;
\]
and
\[
R(z)=\left(\frac{1}{2}-\frac{1+|z|^2}{|1-z|^2}\right)\,
\ln ^2{|z|^2}-\frac{1-|z|^2}{2|1-z|^2}\,
\ln {|z|^2}\,\ln \frac{|1-z|^4}{|z|^2}
\]
\be
+\left(\frac{1}{1-z}-\frac{1}{1-z^*}\right)(z-z^*)\,\int_0^1\frac{dx}{|x-z|^2}\,
\ln \frac{|z|^2}{x^2}\,. \label{R conformal}
\ee
Above we used the complex notations $r=x+iy$ and $r^*=x-iy$ for the two-dimensional
vectors $\vec r= (x,  y)$. Vice versa, two complex numbers $z$ and $z^*$ are equivalent
to the vector $\vec{z}$ with the components $(z+z^*)/2$ and $(z-z^*)/(2i)$.
Furthermore,
$d\vec r =dxdy \equiv drdr^*/2\,,\;\;\delta^{(2)}(\x)=2\delta(r)\delta(r^*)\, $ and
we define $\delta^{(2)}(z)$ in such a way that $\delta^{(2)}(z)=\delta(z)\delta(z^*)/2
=\delta^{(2)}(\vec{z})$.

Note that $K(z)=K(1/z)$.  This property is evident for $K^B(z)$ and for the term
with  $\zeta(3)$. In the case of $R(z)$ it can be proved using the equality
\be
\int_0^1\frac{dx}{|x-z|^2}\,
\ln \frac{|z|^2}{x^2}\,
=\,\ \frac{1}{z-z^*}\left(2\int_0^1\frac{dx}{x}\,
\ln \frac{1-xz^*}{1-xz} -\ln|z|^2\ln \frac{1-z^*}{1-z} \right)
\ee
and the relation
\be
{\rm Li}_2(z)=-{\rm Li}_2\left(\frac{1}{z}\right)-\zeta(2)
-\frac{1}{2}\ln^2\left(-z\right)\,.
\ee

The function $K(z)$ can be expanded in series over the complete set of functions
\be
f_{\nu n}(z)=\frac{1}{\sqrt{2\pi^2}}|z|^{2i\nu}e^{in\phi}\,,\;\;\;z=|z|e^{i\phi}\;,
\ee
with the orthogonality properties
\be
\int \frac{d^2 z}{|z|^2}\, f^{*}_{\mu m}(z)\, f_{\nu n}(z)
=\delta (\mu -\nu )\,\delta_{mn}\,.
\ee
This expansion looks as follows:
\be
K(z)  = \sum_{n=-\infty}^{+\infty}\int_{-\infty}^{+\infty}d\nu \,\omega(\nu, n)
f_{\nu n}(z)\,,
\ee
where the eigenvalues $\omega(\nu, n)$ of the kernel are given  by
\be
\int \frac{d^2 z}{|z|^2}\,K(z) f^*_{\nu n}(z) = \omega(\nu, n)\,.
\ee
They were calculated in~\cite{Fadin:2011we}. It was mentioned already that for the
conformal invariant kernel the eigenvalues do not depend on the total momentum transfer.
Therefore, the eigenvalues are known also for an arbitrary momentum transfer. As it is
well known, an operator is completely defined by its eigenvalues and eigenfunctions.
Formally, one can write
\[
\hat{\cal K}=\sum_n\lambda_n|n\rangle\langle n|,
\]
where $\lambda_n$ are the eigenvalues and $|n\rangle$ are the eigenfunctions normalized
as
\[
\langle n|n'\rangle =\delta_{nn'}\;.
\]
Since we know that the eigenfunctions of the conformal invariant
kernel~\cite{Fadin:2011we} are
\be
\phi _{\nu n}(  q_1,  q_2)=\frac{1}{\sqrt{2\pi^2}}\left(\frac{q_1}{q_2}\right)^{\frac{n}
{2}+i\nu}\left(\frac{q^*_1}{q^*_2}\right)
^{-\frac{n}{2}+i\nu}\,, \;\;\; q_2=q-q_1\,, \label{conformal eigenfunctions}
\ee
with the normalization
\be
\int \frac{{\qs}d\q_1}{\qs_1\qs_2}\left(\phi _{\nu n}(  q_1,  q_2)\right)^*\phi _{\mu m}
(q_1,q_2) =\int \frac{d^2 z}{|z|^2}\, f^{*}_{\mu m}(z)\, f_{\nu n}(z)
=\delta (\mu -\nu )\,\delta_{mn}\,,
\ee
then, denoting the conformal kernel $\hat{\cal K}_c$, we can present it as follows:
\be
K_c(\vec q_1, \vec q_1^{\;\prime}; \vec q)=\sum_{n=-\infty}^{n=+\infty}\int
{d\nu}\;\omega(\nu, n)\phi _{\nu n}\,(  q_1,  q_2)\left(\phi _{\nu n}(q'_1,q'_2)\right)^*
\,.\label{K c as sum and integral}
\ee
But in fact there is no need to calculate this complicated expression. The matter is
that, due to the M\"{o}bius invariance, the kernel $K_c(\vec q_1, \vec q_1^{\;\prime};
\vec q)$ can be written as $K(z)$  given in Eq.~(\ref{K(z)})
with the argument $z=q_1q'_2/(q_2q'_1)$.

Furthermore, if we denote
\be
K(\vec q_1, \vec q_1^{\;\prime}; \vec q)- K_c(\vec q_1, \vec q_1^{\;\prime}; \vec q)=
\frac{\alpha_s^2N_c^2}{32\pi^3}\Delta(\vec q_1, \vec q_1^{\;\prime}; \vec q)\,,
\label{Delta definition}
\ee
then, using the conformal symmetry of $K^B$ and the term with $\zeta(3)$
in Eq.~(\ref{modified kernel}), one obtains from Eqs.~(\ref{modified kernel})
and~(\ref{K(z)})
\be
\Delta(\vec q_1, \vec q_1^{\;\prime}; \vec q) = R(\vec q_1, \vec q_1^{\;\prime}; \vec q) - R(z)\,,
\ee
where $R(\vec q_1, \vec q_1^{\;\prime}; \vec q)$ is given in Eq.~(\ref{R standard}) and
$R(z)$ in Eq.~(\ref{R conformal}) with $z=q_1q'_2/(q_2q'_1)$. Since
$R(\vec q_1, \vec q_1^{\;\prime}; \vec q)$ is not conformal invariant, it cannot be
written using the single variable $z$. Let us define the variables
$z_i=q_1/q'_1\,,\; i=1,2\,,\; z_1/z_2 =z$. Then from Eqs.~(\ref{R standard})
and~(\ref{R conformal}) one has
\[
\Delta(\vec q_1, \vec q_1^{\;\prime}; \vec q) =2\ln\left(\frac{|1-z|^2|z_1|}{|1-z_1|^2|z|}\right)\ln\left(\frac{|1-z|^2|z_2|}
{|1-z_2|^2|z|}\right) +6\ln|z_1|\ln|z_2|-8\frac{1+|z|^2}{|1-z|^2}\ln|z_1|\ln|z_2|
\]
\[
+2\frac{1-|z|^2}{|1-z|^2}\left( \ln|z_1| \ln\frac{|z_1|}{|1-z_1|^2}-\ln|z_2|
\ln\frac{|z_2|}
{|1-z_2|^2} -\ln|z|\ln\frac{|z|}{|1-z|^2}\right)
\]
\[
+2\frac{z-z^*}{|1-z|^2}\biggl[
{\rm Li}_2(z_1)-{\rm Li}_2(z_1^*)-{\rm Li}_2(z_2)+{\rm Li}_2(z_2^*) -{\rm Li}_2(z)
+{\rm Li}_2(z^*)
\]
\be
+\ln|z_1| \ln\frac{1-z_1}{1-z^*_1}-\ln|z_2| \ln\frac{1-z_2}{1-z^*_2}
-\ln|z| \ln\frac{1-z}{1-z^*} \biggr]\,. \label{Delta first}
\ee
Note that $\Delta$ is symmetric with respect to the exchange $1\leftrightarrow 2$, i.e.
$z_1\leftrightarrow z_2, \;\; z\leftrightarrow 1/z$.

The dilogarithms entering Eq.~(\ref{Delta first}) are not independent. Their number can
be reduced using the relation
\be
{\rm Li}_2(z_1/z_2)={\rm Li}_2(z_1)+{\rm Li}_2(1/z_2)+{\rm Li}_2\left(\frac{z_1-1}{z_2-1}\right)+
{\rm Li}_2\left(\frac{z_1(z_2-1)}{z_2(z_1-1)}\right)+\frac12 \ln^2\left(\frac{(z_2-1)}
{z_2(1-z_1)}\right)\,.
\ee
This  gives
\[
\Delta(\vec q_1, \vec q_1^{\;\prime}; \vec q) =2\ln\left(\frac{|1-z|^2|z_1|}{|1-z_1|^2|z|}\right)\ln\left(\frac{|1-z|^2|z_2|}
{|1-z_2|^2|z|}\right) +6\ln|z_1|\ln|z_2|-8\frac{1+|z|^2}{|1-z|^2}\ln|z_1|\ln|z_2|
\]
\[
+2\frac{1-|z|^2}{|1-z|^2}\left( \ln|z_1| \ln\frac{|z_1|}{|1-z_1|^2}-\ln|z_2| \ln\frac{|z_2|}
{|1-z_2|^2} -\ln|z|\ln\frac{|z|}{|1-z|^2}\right)
\]
\[
+2\frac{z-z^*}{|1-z|^2}\biggl[
-{\rm Li}_2\left(\frac{(1-z_1)}{(1-z_2)}\right)+{\rm Li}^*_2\left(\frac{(1-z_1)}{(1-z_2)}\right)-
{\rm Li}_2\left(\frac{(1-z_2)z}{(1-z_1)}\right)+{\rm Li}^*_2\left(\frac{(1-z_2)z}{(1-z_1)}\right)
\]
\be
-\ln\left|\frac{(1-z_1)}{(1-z_2)}\right| \ln\left(\frac{(1-z)z_2(1-z_2^*)}{(1-z^*)z^*_2(1-z_2)}\right)-
\ln\left|\frac{(1-z_2)z}{(1-z_1)}\right| \ln\left(\frac{(1-z)(1-z_1^*)}{(1-z^*)(1-z_1)}
\right)\biggr]\,. \label{Delta through z1 z2}
\ee
Note that
\be
\frac{(1-z_1)}{(1-z_2)}=-\frac{q'_2}{q'_1}\,,\;\; \frac{(1-z_2)z}{(1-z_1)}
=-\frac{q_1}{q_2}\,,\;\;\frac{(1-z)z_2}{(1-z_2)}=-\frac{q}{q'_1}\,,\;\;
\frac{(1-z)}{(1-z_1)}=\frac{q}{q_2}\,.
\ee
The symmetry  with respect to the exchange $1\leftrightarrow 2$, i.e.
$z_1\leftrightarrow z_2, \;\; z\leftrightarrow 1/z$, holds also  here, although it
is not so obvious as before.

\section{Similarity  transformation to the conformal form}

If the kernels $\hat{\cal K}$  and $\hat{\cal K}_c$ coincide in the leading order and
are related by a similarity transformation, there  must  exist an operator $\hat{\cal O}$
satisfying  the commutation relation
\be
[\hat{\cal K}^B, \hat O] =\left(\frac{\alpha_s}{2\pi}\right)^2\frac{1}{8\pi}
\hat{\Delta}\,.\label{Delta as difference}
\ee
One can find a formal expression for this operator allowing to construct the similarity
transformation in perturbation theory.
Indeed, it is enough to calculate the matrix element of the above commutation relation
between the eigenfunctions~(\ref{conformal eigenfunctions}) of the Born kernel with the
corresponding eigenvalues $\omega_{\nu n}^{B}$ in the form
\be
\left(  \omega_{\nu' n'}^{B}-\omega_{\nu n}^{B}\right)  \langle \nu' n'
|\hat{\cal O}|\nu n\rangle=\left(\frac{\alpha_s}{2\pi}\right)^2\frac{1}{8\pi}\langle\nu' n'|\hat{\Delta}|\nu n\rangle. \label{equation for O}
\ee
It can be seen from this equation that the solution $\hat{\cal O}$ exists only if
the operator $\hat{\Delta}$ has vanishing matrix elements between states with the
same eigenvalues. In this case, using the completeness of the functions $|\nu n\rangle$,
we obtain
\be
\hat{{O}}=\left(\frac{\alpha_s}{2\pi}\right)^2\frac{1}{8\pi}\sum_{ n ,  n'}\int d\nu\int d\nu'\frac{|\nu' n'\rangle\langle\nu' n'|
\hat{\Delta}|\nu n\rangle\langle\nu n|}{\omega_{\nu' n'}^{B}-\omega_{\nu n}^{B}}%
\ee
and
\be
\langle \q_{1}, \q_{2}|\hat{\cal O}|\qp_1, \qp_{2}\rangle=
\left(\frac{\alpha_s}{2\pi}\right)^2\frac{1}{8\pi}\sum_{ n ,  n'}\int d\nu\int d\nu'\frac{\langle \q_{1}, \q_{2}|\nu' n'\rangle\langle\nu' n'|
\hat{\Delta}|\nu n\rangle\langle\nu n|\qp_1, \qp_{2}\rangle}{{\omega_{\nu' n'}^{B}-\omega_{\nu n}^{B}}}\;. \label{O matrix elements}
\ee
Since  the kernel $\hat{\Delta}$ is
known in the momentum space (see (\ref{Delta first})), we can transform it into the $(n, \nu)$ representation,
\be
\langle\nu n |\hat{\Delta}|\nu' n' \rangle=\int \frac{\qs d\vec{q}_{1}}{\qs_1(\q-\q_1)^2}
\int \frac{\qs d\vec{q}^{\;\prime}_{1}}{\qps_1(\q-\q_1^{\;\prime})^2}
\langle \nu' n' |\vec{q}%
_{1}^{\;\prime},\vec{q}_{2}^{\;\prime}\rangle\Delta(\q_1, \vec{q}^{\;\prime}_{1}; \q)
\langle\vec{q}_{1},\vec{q}_{2}|\nu n \rangle \label{Delta in nu n}
\ee
using the known eigenfunctions in the momentum space~(\ref{conformal eigenfunctions}),
which allows to find the matrix element
$\langle\vec{q}_{1},\vec{q}_{2}|\hat{\cal O}|\vec{q}_{1}^{\;\prime},
\vec{q}_{2}^{\;\prime}\rangle$.

The eigenfunctions~(\ref{conformal eigenfunctions}) entering
Eq.~(\ref{Delta in nu n})
depend on $r =q_1/q_2$ and $r' =q'_1/q'_2$ ; therefore, it is convenient to express
$\Delta(\q_1, \vec{q}^{\;\prime}_{1}; \q)$  given in Eq.~(\ref{Delta through z1 z2})
in terms of $r, \;r' $ and $z=r/r'\; $. In these variables we have
\[
\Delta \!=\!2\ln^2\frac{|1+r|^2}{|r|} +2\ln^2\frac{|1+r'|^2}{|r'|} -2\ln\frac{|1+r|^2}{|r|}
\ln\frac{|1+r'|^2}{|r'|} -2\ln^2|r| -2\ln^2|r'|+2\ln|r|\ln|r'|
\]
\[
-2\frac{1+|z|^2}{|1-z|^2}\left(\ln^2\frac{|1+r|^2}{|r|} +\ln^2\frac{|1+r'|^2}{|r'|}
-2\ln\frac{|1+r|^2}{|r|}\ln\frac{|1+r'|^2}{|r'|}-\ln^2|z|\right)
\]
\[
+2\frac{1-|z|^2}{|1-z|^2}\left( \ln|r|\ln\frac{|1+r'|^2}{|r'|}
-\ln|r'|\ln\frac{|1+r|^2}{|r|}\right)
\]
\[
+2\frac{z-z^*}{|1-z|^2}\biggl[
-{\rm Li}_2(-r)+{\rm Li}^*_2(-r)+{\rm Li}_2(-r')-{\rm Li}^*_2(-r')
\]
\be
-\ln|r| \ln\left(\frac{1+r}{1+r^*}\right)+\ln|r'| \ln\left(\frac{1+r'}{1+r^{'*}}\right)\biggr]\,. \label{Delta through r r'}
\ee
In terms of the variables $r, r'$ the symmetry with respect to the exchange
$1\leftrightarrow 2$ is equivalent to the symmetry of the above expression  under
the transformations
$r\leftrightarrow 1/r, \;\; r'\leftrightarrow 1/r', \;\;z\leftrightarrow 1/z $.

Note that
\[
2\ln^2\frac{|1+r|^2}{|r|} +2\ln^2\frac{|1+r'|^2}{|r'|} -2\ln\frac{|1+r|^2}{|r|}
\ln\frac{|1+r'|^2}{|r'|} -2\ln^2|r| +2\ln^2|r'|-2\ln|r|\ln|r'|
\]
\be
=2\ln\frac{|1+r|^2}{|1+r'|}\ln\frac{|1+r|^2}{|r|^2}+2\ln\frac{|1+r'|^2}{|1+r|}
\ln\frac{|1+r'|^2}{|r'|^2}\,,
\ee
which demonstrates the absence of singularities at $r=0$. Analogously,
\be
\ln^2\frac{|1+r|^2}{|r|} +\ln^2\frac{|1+r'|^2}{|r'|}
-2\ln\frac{|1+r|^2}{|r|}\ln\frac{|1+r'|^2}{|r'|}-\ln^2|z|=\ln\frac{|1+r|^2}{|1+r'|^2}
\ln\frac{|1+r|^2r^{'2}}{|1+r'|^2r^2}\,.
\ee

The calculations which must be performed in order to obtain $\langle \q_{1},
 \q_{2}|\hat{\cal O}|\qp_1, \qp_{2}\rangle$ by this method are rather complicated. The
first difficulty appears in calculating the matrix elements
$\langle \nu' n' |\hat{\Delta}|\nu n \rangle$.
Here the main problem arises from the term  in $\Delta$ proportional to the expression
\[\frac{1+|z|^2}{|1-z|^2}
\ln\frac{|1+r|^2}{|r|}\ln\frac{|1+r'|^2}{|r'|}\;,
\]
because the corresponding integral is not factorized. It leads to infinite double sums
over poles in the complex planes with positions depending on $n, \nu$ and $n', \nu'$.
At the end one should calculate the complicated
double sum and double integral in Eq.~(\ref{O matrix elements}). But the final result
turns out to be quite simple. Moreover, it can be guessed, as it is shown in the next
section.

\section{Explicit form of the similarity transformation }

To diminish the search area we have to use all possible restrictions implied on this
operator. Important restrictions follow from symmetries and the gauge invariance of
$\hat{\Delta}$ and $\hat{\cal K}_r^B$. Due to the symmetries of $\hat{\Delta}$ and
$\hat{\cal K}^B$ we should look for $\hat{\cal O}$ among operators symmetric under the
interchange $1 \leftrightarrow 2$ and antisymmetric under  transposition. The last
property excludes diagonal terms (proportional to $\delta(\q_1-\qp_1)$ in momentum
space). The non-diagonal part can be taken gauge invariant.

There is only one possibility (up to a coefficient) for such operator without logarithms,
and it is just $\hat{\cal{K}}^B_r$. However, the operator which we are looking for must
contain logarithms, as it follows from Eqs.~(\ref{equation for O})
and~(\ref{Delta through r r'}).   This  enlarges the number of such operators
drastically. But there is one additional argument. It seems quite natural that the
conformal kernel can be obtained by modification of the subtraction procedure used in
the definition of the standard kernel~\cite{Fadin:1998fv}  to separate  leading
and next-to-leading contributions. In this case  the operator $\hat{\cal O}$ must be
proportional to $\hat{\cal{K}}^B_r$. Using the requirement of antisymmetry
under transposition, we come to the conclusion that the most attractive candidate for
$\hat{\cal O}$ is
\be
\hat{\cal O} =C\left[\ln\left(\hat{\vec{q}}_{1}^{\;2}\hat{\vec{q}}_{2}^{\;2}\right),
\hat{\cal{K}}^B_r\right]\,,\label{O t 1}
\ee
where $C$ is some coefficient.

Let us show that indeed the required operator has the form
(\ref{O t 1}) with $C=1/4$.  In the momentum space it looks as
\be
O(\vec q_1, \vec q_1^{\;\prime}; \vec q) \ = \ \frac{\alpha_s N_c}{16\pi^2}\left(  \frac{\vec{q}_{1}^{\;2}\vec
{q}_{2}^{\;\prime\;2}+\vec{q}_{1}^{\;\prime\;2}\vec{q}_{2}^{\;2}}{\vec
{k}^{\;2}}-\vec{q}^{\;2}\right) \ln\left(\frac{{\vec{q}}_{1}^{\;2}{\vec{q}}_{2}^{\;2}}
{\qps_1\qps_2}\right)\,.\label{O t momentum}
\ee
To obtain $\Delta(\vec q_1, \vec q_1^{\;\prime}; \vec q)$ in vector notation
from Eq.~(\ref{Delta through r r'}) the following relations are useful:
\[
2\biggl[
-{\rm Li}_2(-r)+{\rm Li}^*_2(-r)+{\rm Li}_2(-r')-{\rm Li}^*_2(-r')
-\ln|r| \ln\left(\frac{1+r}{1+r^*}\right)+\ln|r'| \ln\left(\frac{1+r'}{1+r^{'*}}
\right)\biggr]\,
\]
\be
= (q_1q_2^{*}-q^{*}_1q_2)I_{\q_1,\q_2} + (q'_2q_1^{'*}-q^{'*}_2q'_1)I_{\qp_1,\qp_2}\,,
\;\; ab^*-a^*b =-2i[\vec{a}\times \vec{b}]\,,\;\;
\ee
\[
\frac{z-z^*}{|1-z|^2}=\frac{2i}{\vks\qs}\left(\vks[\q_1\times\q_2]
-\qs_1[\vk\times\q_2]-\qs_2[\vk\times\q_1]\right)
\]
\be
=\frac{-2i}{\vks\qs}\left(\vks[\qp_1\times\qp_2]
+\qps_1[\vk\times\qp_2]+\qps_2[\vk\times\qp_1]\right)\,.\;\;
\ee
The last equality follows from antisymmetry with respect to $\q_i\leftrightarrow -\qp_i$.

Using these relations, it is easy to obtain
\[
\Delta(\vec q_1, \vec q_1^{\;\prime}; \vec q) =\ln\frac{\qs_1}{\qs}\ln\frac{\qs_2}{\qs}+ \ln\frac{\qps_1}{\qs}\ln\frac{\qps_2}{\qs}
+\ln\frac{\qs_1}{\qps_1}\ln\frac{\qs_2}{\qps_2} -2\frac{\qs_1\qps_2+\qs_2\qps_1}{\vks\qs}
\ln\frac{\qs_1}{\qps_1}\ln\frac{\qs_2}{\qps_2}
\]
\[
+\frac{\qs_1\qps_2-\qs_2\qps_1}{\vks\qs}\left(\ln\frac{\qs_1}{\qs}\ln\frac{\qps_2}{\qs} -
\ln\frac{\qs_2}{\qs}\ln\frac{\qps_1}{\qs}  \right)
\]
\be
+\frac{4}{\qs\vks}\left(\vks[\q_1\times\q_2]
-\qs_1[\vk\times\q_2]-\qs_2[\vk\times\q_1] \right)\left([\q_1\times\q_2]I_{\q_1, \q_2}
-[\qp_1\times\qp_2]I_{\qp_1, \qp_2}  \right)\,.\label{Delta momentum}
\ee
Important properties of $\Delta$  are its symmetries with respect to the exchanges
$\q_1\leftrightarrow -\q_2\,,$ $\,\qp_1\leftrightarrow -\qp_2\,$ and $\q_i
\leftrightarrow -\q_i\,$, as well as the gauge invariance (vanishing at zero momentum
of each reggeon),  which  are easily seen from this representation.

Since the kernel $\hat{\cal K}^B$ contains real and virtual parts, the commutator
\be
\hat{\cal D}=[\hat{\cal K}^B, \hat{\cal O}] =\frac14\left[\hat{\cal K}^B, \left[\ln\left(\hat{\vec{q}}_{1}^{\;2}\hat{\vec{q}}_{2}^{\;2}\right),
\hat{\cal{K}}^B_r\right]\right] \label{D as commutator}
\ee
is naturally separated into two pieces. One is  $\hat{\cal D}_v = [\hat{\omega}_{g_1}+\hat{\omega}_{g_2}, \hat{\cal O}]$ and
gives in the momentum space
\[
D_v(\q_1, \q_2;\;\vk) \ = \ \left(\omega_g(-\vec q_1^{\;2})+\omega_g(-\vec q_2^{\;2})-\omega_g(-\vec q_1^{\;\prime\;2})-
\omega_g(-\vec q_2^{\;\prime\;2})\right)O(\vec q_1, \vec q_1^{\;\prime}; \vec q)
\]
\be
= \
-\frac{\alpha^2_s N^2_c}{32\pi^3}\left(  \frac{\vec{q}_{1}^{\;2}\vec
{q}_{2}^{\;\prime\;2}+\vec{q}_{1}^{\;\prime\;2}\vec{q}_{2}^{\;2}}{\vec
{k}^{\;2}}-\vec{q}^{\;2}\right)\ln^2\left(\frac{{\vec{q}}_{1}^{\;2}{\vec{q}}_{2}^{\;2}}
{\qps_1\qps_2}\right)\,. \label{D v}
\ee
The  piece $\hat{\cal D}_r=[\hat{\cal K}_r^B,\hat{\cal O}]$ can be written in the
momentum space as
\[
{D}_r(\q_1, \q_2;\;\vk) = \frac{\qs_1\qs_2}{4}\int{d\p}\,\frac{K_r(\q_1, \q_2;\;\q_1-\p)}
{\qs_1\qs_2}
\frac{K_r(\p, \q-\p; \;\p-\qp_1)}{\ps(\q-\p)^2}
\]
\be
\times \ln\left(\frac{(\ps)^2}{\qs_1\qps_1}\frac{((\q-\p)^2)^2}{\qs_2
\qps_2}\right)\,.\label{real integral}
\ee
A convenient way to calculate this integral is to use complex variables for the
two-dimensional vectors and to perform the pole expansion of the integrand. We have
\[
\frac{K_r(\q_1, \q_2;\;\q_1-\p)}{\qs_1\qs_2}
\frac{K_r(\p, \q-\p; \;\p-\qp_1)}{\ps(\q-\p)^2} \
\]
\[
= \  \left(\frac{\alpha_s N_c}{4\pi^2}
\right)^2
\biggl[\left(\frac{1}{(\qp_1-\p)^2}+
\frac{1}{(\q_1-\p)^2} \right)\left(  \frac{\vec{q}_{1}^{\;2}\vec
{q}_{2}^{\;\prime\;2}+\vec{q}_{1}^{\;\prime\;2}\vec{q}_{2}^{\;2}}{\vec
{k}^{\;2}\vec{q}_{1}^{\;2}\vec
{q}_{2}^{\;2}}-\frac{\vec{q}^{\;2}}{\vec{q}_{1}^{\;2}\vec
{q}_{2}^{\;2}}\right)
\]
\[
 -\frac{1}{(q_1-p)(q'^*_1-p^*)}\frac{q'_1q^*}{\qs_1 k q^*_2}-\frac{1}{(q_1-p)p^*}
\frac{q'_2q^*}{\qs_2 k q^*_1} + \frac{1}{(q'_1-p)(q^*-p^*)}\frac{q'_2q^*}{\qs_2 k q^*_1}
\]
\be
+ \frac{1}{(q'_1-p)p^*}\frac{q'_1q^*}{\qs_1 k q^*_2}-\frac{1}{(q_1-p)(q'^*_1-p^*)}
\frac{1}{\vks}\left|\frac{q'_1}{q_1}+\frac{q'_2}{q_2}\right|^2+ \frac{1}{p(q^*-p^*)}\frac{\qs}{\qs_2
 \qs_2} +{\rm c.c.}\biggr]\,.
\ee
Taken separately, each term in this expansion gives an ultraviolet-divergent integral.
Of course, the divergences cancel in their sum. Introducing the intermediate cutoff
$\ps\leq \Lambda^2$, one has
\[
\int\frac{d\p}{\pi}\ln\left(\frac{(\ps)^2}{\qs_1\qps_1}\right)
 \left(\frac{1}{(\qp_1-\p)^2}+\frac{1}{(\q_1-\p)^2} \right)\theta(\Lambda^2-\vec p^{\;2}) =\ln^2\left(\frac{\Lambda^2}
 {\qs_1}\right)+\ln^2\left(\frac{\Lambda^2}
 {\qps_1}\right)\,,
\]
\[
\int\frac{d \vec p}{\pi}\frac{2}{(a- p)(b^*- p^*)}
\ln\left(\frac{\vec p^{\;2}}{\mu^2}\right)
\theta(\Lambda^2-\vec p^{\;2})=
\ln\left(\frac{\Lambda^2}{(\vec a -\vec b)^{2}}\right)
\ln\left(\frac{\Lambda^2(\vec a -\vec b)^{2}}{\mu^4}\right)
\]
\begin{equation}
+\ln\left(\frac{(\vec a -\vec b)^{2}}{\vec b^{\;2}}\right)
\ln\left(\frac{(\vec a -\vec b)^{2}}{\vec a^{\;2}}\right)
+(ab^*-a^*b)I_{\vec a,-\vec b }~. \label{int master}
\end{equation}
Using also equalities
\[
\frac{q'_1q^*}{\qs_1 k q^*_2}+{\rm c.c.}
=\frac{\qs_1\qps_2-\qs_2\qps_1-\qs\vks}{\qs_1\qs_2\vks}\;,
\]
\[
\frac{q'_2q^*}{\qs_2 k q^*_1}+{\rm c.c.}
=\frac{\qs_1\qps_2-\qs_2\qps_1+\qs\vks}{\qs_1\qs_2\vks}\,,
\]
\be
\frac{1}{\vks}\left|\frac{q'_1}{q_1}+\frac{q'_2}{q_2}\right|^2=
\frac{2(\qs_1\qps_2+\qs_2\qps_1)-\qs\vks}{\qs_1\qs_2\vks}\,,
\ee
we obtain
\[
D_r(\q_1, \q_2;\;\vk) = \frac{\alpha^2_s N^2_c}{32\pi^3}  \qs\biggl[\ln\frac{\qs_1}{\qs}\ln\frac{\qs_2}{\qs}+ \ln\frac{\qps_1}{\qs}\ln\frac{\qps_2}{\qs}
-\ln\frac{\qs_1}{\qps_1}\ln\frac{\qs_2}{\qps_2}
\]
\[
+\left(\frac{\qs_1\qps_2+\qs_2\qps_1}{\vks\qs}-1\right)\left(
\ln^2\frac{\qs_1}{\qps_1}+\ln^2\frac{\qs_2}{\qps_2}\right)
\]
\[
+\frac{\qs_1\qps_2-\qs_2\qps_1}{\vks\qs}\left(\ln\frac{\qs_1}{\qs}\ln\frac{\qps_2}{\qs} -
\ln\frac{\qs_2}{\qs}\ln\frac{\qps_1}{\qs}  \right)
\]
\be
+\frac{4}{\qs\vks}\left(\vks[\q_1\times\q_2]
-\qs_1[\vk\times\q_2]-\qs_2[\vk\times\q_1] \right)\left([\q_1\times\q_2]I_{\q_1, \q_2}
-[\qp_1\times\qp_2]I_{\qp_1, \qp_2}  \right)\biggr]\,. \label{D r}
\ee
The total commutator $\hat{\cal D} = [\hat{\cal K}^B, \hat{\cal O}]$ is defined in the
momentum space by the sum of the two pieces given in Eqs.~(\ref{D v})
and~(\ref{D r}):
\[
D(\q_1, \q_2;\;\vk) = \frac{\alpha^2_s N^2_c}{32\pi^3}
 \qs\biggl[\ln\frac{\qs_1}{\qs}\ln\frac{\qs_2}{\qs}+ \ln\frac{\qps_1}{\qs}\ln\frac{\qps_2}{\qs}
+\ln\frac{\qs_1}{\qps_1}\ln\frac{\qs_2}{\qps_2}
\]
\[
-2\frac{\qs_1\qps_2+\qs_2\qps_1}{\vks\qs}
\ln\frac{\qs_1}{\qps_1}\ln\frac{\qs_2}{\qps_2}+\frac{\qs_1\qps_2-\qs_2\qps_1}{\vks\qs}\left(\ln\frac{\qs_1}{\qs}\ln\frac{\qps_2}{\qs} -
\ln\frac{\qs_2}{\qs}\ln\frac{\qps_1}{\qs}  \right)
\]
\be
+\frac{4}{\qs\vks}\left(\vks[\q_1\times\q_2]
-\qs_1[\vk\times\q_2]-\qs_2[\vk\times\q_1] \right)\left([\q_1\times\q_2]I_{\q_1, \q_2}
-[\qp_1\times\qp_2]I_{\qp_1, \qp_2}  \right)\biggr]\,. \label{D}
\ee
Comparing Eq.~(\ref{D}) with Eq.~(\ref{Delta momentum}) and taking into account
Eq.~(\ref{D as commutator}), one sees that indeed Eq. (\ref{Delta as difference}) is fulfilled, if  $\hat{\cal O}$ is given by
(\ref{O t 1}) with $C=1/4$. Using (\ref{Delta definition}), we conclude that
\be
\hat{\cal K} -\hat{\cal K}_c = \frac14\left[\hat{\cal K}^B, \left[\ln\left(\hat{\vec{q}}_{1}^{\;2}\hat{\vec{q}}_{2}^{\;2}\right),
\hat{\cal{K}}^B_r\right]\right]\,. \label{final}
\ee
It means that indeed conformal and standard forms of the kernel are connected
by a similarity transformation. Moreover, this transformation is equivalent to the
change of the subtraction procedure in the definition of the NLO
kernel~\cite{Fadin:1998fv}.

\section{Conclusion}

In this paper, we have shown that the standard form of the modified BFKL kernel
(i.e. the BFKL kernel in $N=4$ SUSY for the adjoint representation of the gauge
group with subtracted gluon trajectory depending on total momentum transfer) can be
reduced by a similarity transformation to a form which is M\"{o}bius invariant in the
momentum space. The transformation is given by Eq.~(\ref{final}).
This transformation is equivalent to the change of the subtraction procedure
to separate  leading and next-to-leading contributions used in the definition of
the NLO kernel~\cite{Fadin:1998fv}.

The M\"{o}bius  invariant kernel was used for calculation of the NLO remainder function
in~\cite{Fadin:2011we} with the M\"{o}bius invariant convolution of the NLO BFKL impact
factors (which was called for brevity simply impact factor) obtained
in~\cite{Lipatov:2010qg} from direct two-loop calculations and with the energy scale
$s_0$ chosen in such a way that the ratio $s/s_0=1/\sqrt{u_2u_3}=s\q^{\;2}/\sqrt{\q_1^{\;2}\q_2^{\;2}\q_1^{\;\prime \;2}\q_2^{\;\prime \;2}}$  is M\"{o}bius
invariant. In principle, one can use different definitions of $s_0$ with  M\"{o}bius
invariant ratio $s/s_0$. The definition used in~\cite{Fadin:2011we} is natural because
of $t$-channel factorization of the amplitude in the Regge theory and is matched to the
definition of the NLO BFKL kernel and impact factors~\cite{Fadin:1998fv}. But for
complete assurance  of  M\"{o}bius invariance of the remainder function, one
should check that the convolution of the last impact factors is reduced to the impact
factor used in~\cite{Fadin:2011we} by the same similarity transformation as the kernel.

\vspace{0.5cm} {\textbf{{\Large Acknowledgments}}}

\vspace{0.5cm} V.S.F. thanks the Dipartimento di Fisica
dell'Universit\`{a} della Calabria and the Istituto Nazionale di
Fisica Nucleare (INFN), Gruppo Collegato di Cosenza, for warm hospitality
while part of this work was done and for financial support.

\end{document}